\documentclass[pra,twocolumn,showpacs,superscriptaddress]{revtex4}
\usepackage{graphicx}
\usepackage{amsfonts,amsmath,amssymb,float}
\usepackage{bm,dsfont}

\begin{document}
\title{Local quantum control of Heisenberg spin chains}
\author{Rahel Heule}
\affiliation{Department of Physics, University of Basel,
 Klingelbergstrasse 82, CH-4056 Basel, Switzerland}
 
\author{C. Bruder}
\affiliation{Department of Physics, University of Basel,
 Klingelbergstrasse 82, CH-4056 Basel, Switzerland}
 
\author{Daniel Burgarth}
\affiliation{Institute for Mathematical Sciences, Imperial College London, SW7 2PG, United
 Kingdom} 
\affiliation{QOLS, The Blackett Laboratory, Imperial College London, Prince Consort Road,
 SW7 2BW, United Kingdom}

\author{Vladimir M. Stojanovi\'c}
\email{vladimir.stojanovic@unibas.ch}
\affiliation{Department of Physics, University of Basel,
 Klingelbergstrasse 82, CH-4056 Basel, Switzerland} 

\date{\today}

\begin{abstract}
Motivated by some recent results of quantum control theory, 
we discuss the feasibility of local operator control in arrays of
interacting qubits modeled as isotropic Heisenberg spin chains. Acting on 
one of the end spins, we aim at finding piecewise-constant control pulses 
that lead to optimal fidelities for a chosen set of quantum gates. We analyze 
the robustness of the obtained results for the gate fidelities to random errors 
in the control fields, finding that with faster switching between
piecewise-constant controls the system is less susceptible to these
errors. The observed behavior falls into a generic class of physical
phenomena that are related to a competition between resonance- and
relaxation-type behavior, exemplified by motional narrowing in NMR
experiments. Finally, we discuss how the obtained optimal gate
fidelities are altered when the corresponding rapidly-varying
piecewise-constant control fields are smoothened through spectral
filtering.
\end{abstract}

\maketitle

\section{Introduction}

The ability to manipulate the state and/or the dynamics of complex
quantum systems is one of the main objectives of quantum information
science~\cite{NielsenChuangBook}. In this context, quantum
control~\cite{D'AlessandroBook} has been successfully utilized to
address two general classes of issues. In {\em state-selective}
control, the main question is how to steer a physical system from an
initial reference state to a desired final state. In {\em operator 
(state-independent)} control, one seeks to implement a predetermined
unitary transformation (target quantum gate) irrespective of the
initial state of the system, which is often unknown. The rigorous
mathematical foundations of the field, based on the notion of
controllability and framed using Lie-algebraic concepts, have long been
known~\cite{ControlConcepts}. Recent quantum-control studies have been
focused around two sets of questions.

Firstly, even when a system is fully controllable in principle, it is
of interest to know how to control it most efficiently, taking into
account various practical constraints. Issues of this type are not
likely to yield universal answers, except for well-understood general
topological features of optimal-control
landscapes~\cite{RabitzTwoPapers}. 
Secondly, it is desirable to know whether the system can
be partly or fully controlled by acting only on a small subsystem.
This is the main idea behind the {\em local-control} (minimal 
actuation) approach. Such an approach is applicable only in
interacting systems like coupled spin chains, where the interaction
will effectively turn the local control applied to the small subsystem
to a global one. Uses of such ``always-on'' interactions in quantum
information processing include, e.g., systems which can serve as data
buses~\cite{Bose:07} enabling
state-~\cite{Romito+:05,Lyakhov+Bruder,Burgarth:07,Caneva+:09} and
entanglement transfer~\cite{Maruyama+:07}. Several notable results
have recently been
reported~\cite{Schirmer:08,Burgarth+:09,XYendspins:10,Wang+:10}. For
instance, controlling only one of the end spins of an $XXZ$-Heisenberg
spin chain ensures full controllability of the
chain~\cite{Burgarth+:09}. In addition, universal quantum computation
with the $XY$ spin chain can be effected by controlling two spins at the
end of the chain~\cite{XYendspins:10}. Finally, 
a magnetic field in the $z$-direction acting only on a single
spin suffices for generating perfect entanglement between the
ends of a Heisenberg chain~\cite{Wang+:10}.

Apart from being interesting from a conceptual point of view, the
local-control approach also has some practical importance. In addition to 
its easier implementation, lowering the number of control
actuators reduces the effects of decoherence.  This is of crucial
importance in many quantum-computer architectures, such as
superconducting qubits~\cite{Montangero+:07,Fisher+:10}.

In this paper, motivated in part by the rigorous results of
Ref.~\onlinecite{Burgarth+:09}, we consider an implementation of
local operator control in arrays of interacting qubits modeled as 
spin chains with isotropic Heisenberg interaction. 
Acting only on one of the end spins, we determine 
piecewise-constant control fields resulting in the highest possible fidelities
for a chosen set of quantum gates.  By treating the amplitudes of
these control fields as optimization variables, we find the optimal
fidelities for chains with up to four spins~\cite{RahelThesis}.
While our work focuses on Heisenberg-coupled spin chains, its results
are relevant for any physical implementation of interacting qubit
arrays.

While some elements of the quantum control of spin chains have already
been studied~\cite{Schulte+:05} -- even in the open-system
scenario~\cite{Grace+Lapert} -- we discuss in detail some aspects that
have so far not received due attention.  In particular, we carry out a
sensitivity analysis, i.e., consider the robustness of the obtained
results for the gate fidelities with respect to random errors in the
control fields.  Our analysis shows that with faster switching between
piecewise-constant controls, the system is less susceptible to these
errors. Importantly, we explain this behavior by making a link to a
class of phenomena exemplified by motional narrowing in NMR
experiments~\cite{CallaghanBook}. To bridge the gap between
theoretical considerations and future experimental implementations of
the system under study, we discuss how the optimal fidelities are
affected when the control pulses are smoothened by eliminating
high-frequency components in their Fourier spectra (spectral
filtering). We establish some qualitative criteria regarding the
performance of such an approach.

The paper is organized as follows. To set the stage, in
Sec.~\ref{system} we introduce the system and our
control objectives. In Sec.~\ref{resultspin234}
we present detailed results for the optimal gate fidelities and the
corresponding control fields in chains with three and four
spins. We also discuss minimal action times for the control fields
needed to realize certain quantum gates.  Section~\ref{sensitivity} is
set aside for the sensitivity analysis, while 
Sec.~\ref{smoothing} deals with the effects of spectral filtering of
the optimal piecewise-constant control fields on the resulting gate
fidelities. We conclude  with a brief summary of the paper and some
general remarks in Sec.~\ref{sumconclude}. Some numerical details are 
described in the Appendix.

\section{System and method} \label{system}
\subsection{Hamiltonian and control objectives}

We consider an isotropic Heisenberg spin-$1/2$ chain of length $N_s$,
with control fields acting on the first spin only. It is governed by
the total Hamiltonian
\begin{equation}
H(t)=H_{0}+ H_c(t)\:,
\end{equation}
where $H_{0}$ is the Heisenberg Hamiltonian
\begin{equation}\label{H_0}
H_{0}=J\sum_{i=1}^{N_{s}-1}\:
\left(S_{ix}S_{i+1,x}+S_{iy}S_{i+1,y}+S_{iz}S_{i+1,z}\right)\:,
\end{equation}
while  
\begin{equation}
H_c(t)=h_x(t)S_{1x}+h_y(t)S_{1y}
\end{equation}
is the Zeeman-like control Hamiltonian.  The time
dependence of the control fields $h_x(t)$ and $h_y(t)$ will be
specified shortly.  For convenience, throughout this paper we set
$\hbar=1$. As a consequence, all frequencies and control-field
amplitudes can be expressed in units of the coupling strength $J$, and
all times in units of $1/J$.

Regarding the choice of Hamiltonian in Eq.~\eqref{H_0} two remarks are
in order. Firstly, while the algebraic result of
Ref.~\onlinecite{Burgarth+:09} is even applicable to the more general
(anisotropic) $XXZ$-Heisenberg Hamiltonian, for simplicity we here
discuss the isotropic case only. Secondly, whether the spin chain is
ferromagnetic ($J<0$) or antiferromagnetic ($J>0$) is unessential for
our present purposes, because we are concerned with operator control.
For concreteness we assume that $J>0$.

In Ref.~\onlinecite{Burgarth+:09}, using the graph infection
criterion, it was shown that controlling the $x$ and $y$ components of
the field acting on the first spin guarantees the complete
controllability of a Heisenberg chain.  Moreover, there are unitary
transformations that require even smaller degree of control, i.e., a
control field only in one direction. For instance, to achieve a
spin-flip operation $X$ on the last spin of the chain, where $X$ is
the Pauli matrix, one needs only a control field in the $x$
direction. The corresponding dynamical Lie algebra $\mathcal{L}_x$,
generated by $\{-iH_0,-iS_{1x}\}$, is a subalgebra of $su(d)$. The
action of the $N_s$-qubit gate $X_{N_s}$, which flips the last spin in
the Heisenberg chain, is defined by
\begin{equation}
X_{N_s}:=\mathds{1}\otimes\mathds{1}\otimes\hdots\otimes\mathds{1}\otimes X\:.
\end{equation}
To prove the reachability of $X_{N_s}$, i.e., that $X_{N_s}\in
e^{\mathcal{L}_x}$ [where $e^{\mathcal{L}_x}$ is the connected Lie
  subgroup of $SU(d)$ with Lie algebra $\mathcal{L}_x$], we have to
show that there exists an element $A$ of the dynamical Lie algebra
$\mathcal{L}_x$ such that $X_{N_s}=e^{A}$. By calculating the
(repeated) commutators of the operators which generate the algebra, we
find that $X_{N_s}$ is an element of $\mathcal{L}_x$. Since $X_{N_s}$
is both unitary and Hermitian (hence $X_{N_s}^2=\mathds{1}$), the
operator $A=-i\frac{\pi}{2}X_{N_s}$, an element of $\mathcal{L}_x$,
has the property that $e^{A}=-iX_{N_s}$. This concludes the proof that
the $x$ field is sufficient to reach $X_{N_s}$. More generally, any
unitary $U$, which is also Hermitian and for which $-iU$ belongs to
the dynamical Lie algebra $\mathcal{L}$, is an element of
the reachable set $e^\mathcal{L}$.

In the following, our control objective is the realization of concrete
quantum gates.  We discuss gates that require control fields in the
$x$ and $y$ directions, as well as those that entail only an $x$
control field. Apart from spin-flip ($\mathrm{NOT}$) gates, we
implement entangling two-qubit gates such as
\begin{equation} 
\mathrm{CNOT}_{N_s}:=\mathds{1}\otimes\mathds{1}
\otimes\hdots\otimes\mathds{1}\otimes\mathrm{CNOT}\:,
\end{equation} 
which performs the controlled-$\mathrm{NOT}$ ($\mathrm{CNOT}$)
operation on the last two qubits in the chain.  Another example is
square root of $\mathrm{SWAP}$ ($\sqrt{\mathrm{SWAP}}$). In operator
control, the figure of
merit in the realization of these gates is the gate fidelity
\begin{equation}\label{deffidelity}
F(t)=\frac{1}{d}\big|\mathrm{tr}
\big[U^{\dag}(t)U_{\mathrm{target}}\big]\big|\:,
\end{equation}
where $U(t)$ is the time-evolution operator of the system at time $t$
and $U_{\mathrm{target}}$ stands for the desired quantum gate.

\subsection{Time-evolution for piecewise-constant control fields} 
\label{pccontrols}

For many classes of systems, the implementation of complicated
time-dependent potentials is rather difficult. In what follows, we
resort to simple piecewise-constant controls. Importantly, we retain
the full Hilbert space of the system, unlike some previous studies
that make use of the single-excitation subspace.~\cite{Schirmer+:01}
This puts constraints on the system size.

Assume that we want to achieve an arbitrary target unitary at a time
$t=t_f$. At $t=0$ we apply an $x$ control pulse to the first spin of
the chain with amplitude $h_{x,1}$ which is constant throughout the
pulse duration $T$. That is, the system evolves under the action of
the Hamiltonian $H_{x,1}\equiv H_0+h_{x,1}S_{1x}$.  Then we apply a
$y$ control pulse with amplitude $h_{y,1}$ during the second time
interval of length $T$, whereby the system is governed by the
Hamiltonian $H_{y,1}\equiv H_0+h_{y,1}S_{1y}$. This sequence of
alternate $x$ and $y$ control pulses is repeated until $N_t$ pulses
have been completed at the time $t_f\equiv N_tT$. The full
time-evolution operator can be expressed as
\begin{equation}
U(t_f)=U_{y,N_t/2}\cdot U_{x,N_t/2}\cdot...\cdot U_{y,1}\cdot U_{x,1}\:,
\label{time_evolution_xy}
\end{equation}
where $U_{x,i}\equiv e^{-iH_{x,i}T}$ and $U_{y,i}\equiv
e^{-iH_{y,i}T}$ are the respective time-evolution operators
corresponding to $H_{x,i}$ and $H_{y,i}$. 
We evaluate $U_{x,i}$ and $U_{y,i}$ using
their spectral forms. If the desired gate is
achievable by control of the $x$ field only, it is sufficient to apply
in each time interval an $x$ field of variable amplitude.

\section{Three- and four-spin chains} \label{resultspin234}

For the numerical maximization of the fidelity
[Eq.~\eqref{deffidelity}] with respect to the $N_t$ field amplitudes,
we use a quasi-Newton method developed by Broyden, Fletcher, Goldfarb,
and Shanno (BFGS-algorithm)~\cite{NRfortranBook}.  We first choose an
initial guess for the control-field amplitudes. The algorithm then
generates iteratively new sequences of field amplitudes such that at
each iteration point the fidelity is increased, and terminates as soon
as the desired accuracy is reached. This procedure ensures the
convergence to a local maximum, but of course does not guarantee 
the convergence to a globally-optimal sequence of control field amplitudes.

We perform maximizations with varying number $N_t$ and durations $T$
of pulses, and hence different total evolution times $t_f=N_tT$. It is
important to stress that, instead of fixing the pulse durations and
maximizing over the field amplitudes, we could as well keep the
amplitudes constant and treat the pulse durations as free control
parameters~\cite{Schirmer:08,Schirmer+Pemberton:09}. Yet, we choose to
maximize over the field amplitudes as this approach allows us to fix
easily the total evolution time and estimate its minimal value needed to
implement the desired gate. Although controllability implies the
existence of a control that can enable realization of a desired
quantum gate, it does not provide any information about the minimal
time over which this realization is possible.

\begin{figure}[t!]
\includegraphics[width=0.85\linewidth]{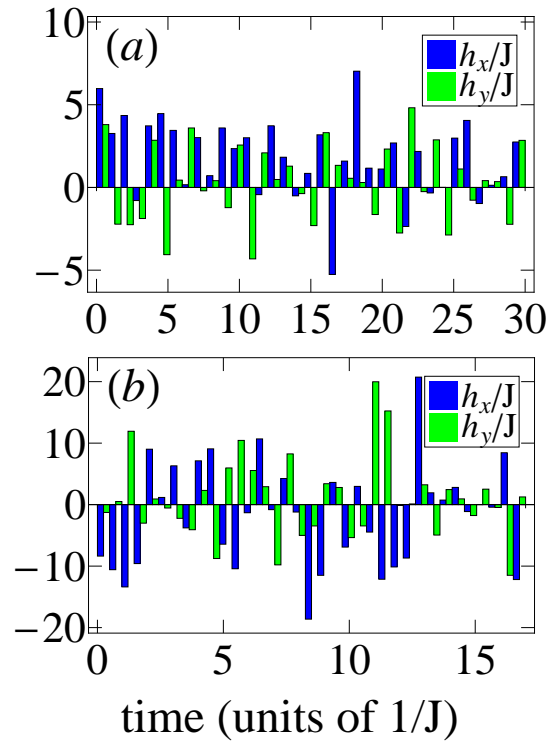}
\caption{\label{optimal_fields}(Color online) Examples of optimal control fields
  for the $X_3$ gate, corresponding to: (a) $N_t=70$ and $T=0.429$,
  with the resulting optimal fidelity of $1-10^{-8}$, and (b) $N_t=70$
  and $T=0.243$ (minimal total time $t_{f}=17$), with the resulting
  optimal fidelity of $1-10^{-6}$.}
\end{figure}

We now give explicit examples of control sequences that implement all
of our chosen gates in chains of three and four spins, among them
entangling gates which are essential for universal quantum
computation.  Two optimal sequences of control pulses implementing the
spin-flip gate $X_{N_s}$ for a three-spin chain ($N_s=3$) with $x$ and
$y$ pulses are shown in Fig.~\ref{optimal_fields} [the optimal field
in Fig.~\ref{optimal_fields}(b) corresponds to the minimal time
needed].  For given total time longer than the minimal time, we can
generate piecewise-constant control fields with fidelities arbitrarily
close to unity by increasing the switching rate. For instance, in the
case of implementing the gate $X_3$ by control of the $x$ and $y$
fields, for a total time of $t_f=30$ and $N_t=10,20,30,40,50,60,70$,
we respectively obtain $F=0.536,0.764,0.904,0.964,0.992,0.999,1-10^{-8}$.

We give here an estimate of the minimal time for the realization of
the gate $X_{N_s}$ (operating on three- and four-spin chains) by
controlling the $x$ and $y$ field. The implementation of this unitary
is possible for $N_s=3$ and $N_s=4$ within the respective times $t_f=17$ 
and $t_f=70$, which are lower bounds on the evolution time needed. Unlike in
Ref.~\onlinecite{XYendspins:10}, the size of the dynamical Lie algrebra
scales exponentially with the number of qubits. We therefore expect
that the minimal time grows rapidly with the chain length, and this is
a limiting factor for extending this control procedure.

For each fixed chain length, the minimal times for realizing the spin-flip 
and CNOT gates are similar. This result seems somewhat related to the main conclusion of
Ref.~\onlinecite{Ghosh+Geller:10}, where the $\mathrm{CNOT}$ gate is
implemented for two qubits coupled through a variety of interactions
($XY$, Heisenberg, Ising) and the total gate time does not depend
significantly on the choice of interaction as long as it is not of
pure Ising type.

Recalling that gates like $X_{N_s}$ require only an $x$ control field,
it is interesting to compare the minimal times for their implementation
depending on the degree of control. We find that with control of the
$x$ field only, the respective gates can be achieved in approximately
the same time as with control of both the $x$ and $y$ fields. 

\section{Sensitivity analysis} \label{sensitivity}
\subsection{Preliminary considerations}

In the following, we analyze the sensitivity of the fidelity to random
errors in the control fields. To this end, we add random numbers from
a uniform distribution of half-width $\delta$ to the optimal
control-field amplitudes. For given $\delta$, we generate a few
hundred sequences of random numbers. In this way we obtain a large
sample of $N\sim 1000$ control fields affected by random noise, for
which we recalculate the fidelity. 

\begin{figure}[t!]
\includegraphics[width=0.85\linewidth]{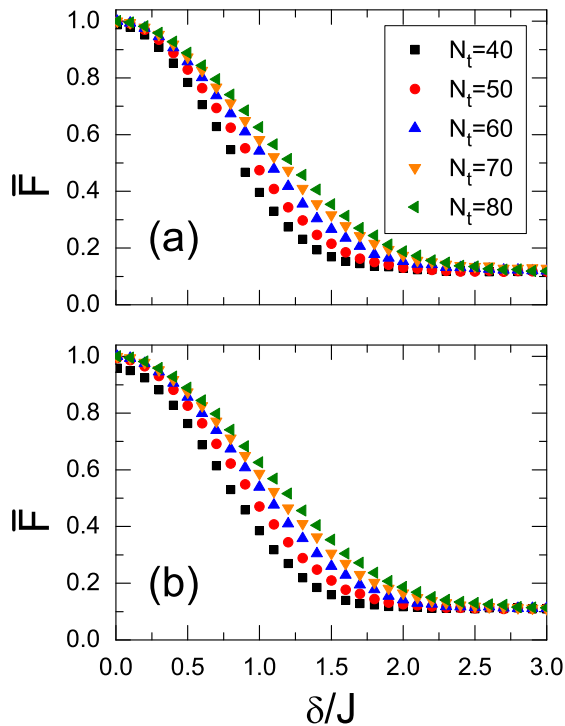}
\caption{\label{sensitivity_hxhy}(Color online) Average fidelity
  versus half-width ($\delta$) for random noise affected optimal
  fields corresponding to (a) $X_3$ gate, and (b) $\mathrm{CNOT_3}$
  gate, both implemented by alternate $x$ and $y$ controls with a
  total evolution time $t_f=30$.}
\end{figure}

\begin{figure}[t!]
\includegraphics[width=0.85\linewidth]{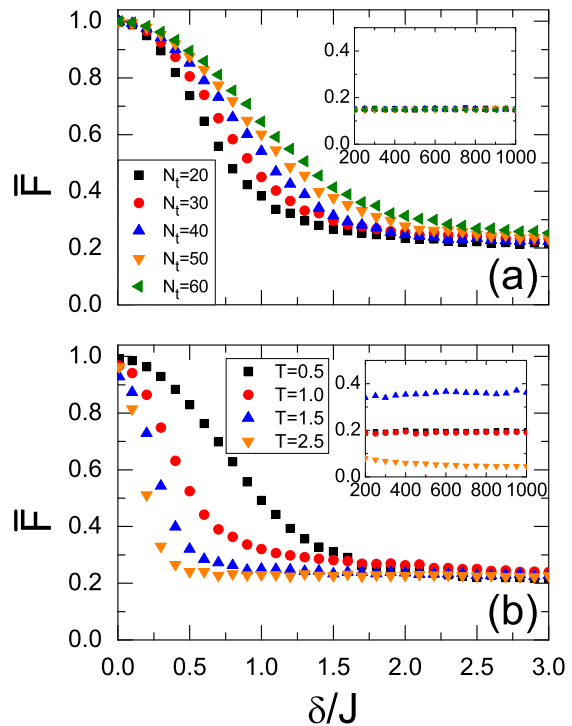}
\caption{\label{sensitivity_hx}(Color online) Average fidelity versus
  half-width ($\delta$) for random noise affected optimal fields
  corresponding to the implementation of the $X_3$ gate by control of
  the $x$ field only: (a) fixed total evolution time ($t_f=25$) and
  (b) fixed number of pulses ($N_t=70$). The insets show the same
  curves for large values of the parameter $\delta$.}
\end{figure}

In the following, we discuss the behavior of the average fidelity 
$\bar{F}=\sum_{i=1}^N F_i/N$, where the $F_i$ are fidelities calculated for 
specific realizations of the random field. We note that the standard deviation
$\sigma_{\scriptscriptstyle F}$ increases with $\delta$: for
$\delta=0.01$ it is found to be of the order of $10^{-5}$, while for
$\delta=1.0$ we obtain $\sigma_{\scriptscriptstyle F}\sim 0.1$.
Figures~\ref{sensitivity_hxhy} and~\ref{sensitivity_hx}
illustrate how the strength of the randomness in the optimal field
amplitudes affects the average fidelity. While
Fig.~\ref{sensitivity_hxhy} shows the results for the gates $X_3$ and
$\mathrm{CNOT}_3$ both realized by a sequence of alternate $x$ and $y$
pulses, Fig.~\ref{sensitivity_hx} refers to the implementation of
$X_3$ by control of the $x$ field only. We see that $\bar{F}$ is less
susceptible to random noise for shorter pulse durations $T$. A
physical understanding of this phenomenon is provided in
Sec.~\ref{nmr}. 

Starting from optimal fields, the shape of the fidelity decay with
increasing $\delta$ is dependent on the number of control pulses $N_t$
and their length $T$, but not on the pulse amplitudes. The saturation 
regime of the average fidelity, indicating full randomization,
sets in for $\delta\gtrsim J$. Importantly, the saturation of the
fidelity is an intrinsic property of our system. If the system is
completely controllable, the saturation value is universal, i.e.,
independent of the concrete shape of the control field, provided the
evolution time $t_f$ and the number of pulses $N_t$ allow the
generation of any unitary (illustrated in Fig.~\ref{sensitivity_hxhy}
for $t_f=30$). In contrast, if the dynamical Lie algebra of the
system is a proper subalgebra of $su(d)$, for different evolution times
$t_f$ the average fidelity may saturate at different values, and the
saturation may require much stronger random fields in comparison to
the completely controllable system (for illustration, see
Fig.~\ref{sensitivity_hx}).

\subsection{Fidelity saturation}

To understand the value of the fidelity at saturation, we invoke the
notion of the average gate fidelity~\cite{AverageFidelity}. If
$\varepsilon$ is a trace-preserving quantum operation pertaining to a
quantum system with $d$-dimensional Hilbert space and $U$ a certain
quantum gate, the average gate fidelity
\begin{equation}\label{avgatefid1}
\bar{F}(\varepsilon,U)=\frac{\sum_{j=1}^{d^2}\mathrm{tr}
\left[UU_j^{\dag}U^{\dag}\varepsilon(U_j)\right]+d^2}{d^2(d+1)}\:,
\end{equation}
where $\{U_j\}$ is an orthonormal unitary-operator basis of the
complex vector space $\mathbb{C}^{d\times d}$, quantifies how well
$\varepsilon$ approximates $U$. If $\varepsilon$ implements the gate
$U$ perfectly, this mapping is given by
$\varepsilon:\rho\rightarrow\varepsilon(\rho)=U\rho U^{\dag}$,
implying that $\bar{F}(\varepsilon,U)=1$; otherwise, the
implementation is noisy and, consequently,
$\bar{F}(\varepsilon,U)<1$. An equivalent form of
Eq.~\eqref{avgatefid1} -- which involves the traceless generators
$\{T_j\}$ of $SU(d)$ [recall that $\mathrm{tr}(T_jT_k)=\delta_{jk}/2$,
see e.g. Ref.~\onlinecite{PfeiferBook}]
reads~\cite{Bagan+:03}:
\begin{equation}\label{avgatefid2}
\bar{F}(\varepsilon,U)=\frac{1}{d}+\frac{2}{d(d+1)}\sum_{j=1}^{d^2-1}
\mathrm{tr}\left[UT_jU^{\dag}\varepsilon(T_j)\right]\:.
\end{equation}
Equations~\eqref{avgatefid1} and~\eqref{avgatefid2} are both derived
by integrating over the uniform Haar measure~\cite{AverageFidelity},
hence requiring a uniform starting distribution of
states and/or operators. Since the quantum control system governed by the
Hamiltonian $H(t)=H_0+h_x(t)S_{1x}+h_y(t)S_{1y}$ allows to generate
any element of $SU(d)$, Eq.~\eqref{avgatefid2}, which is derived
explicitly in terms of the $SU(d)$ group generators, provides a
suitable expression for calculating the average gate fidelity. 
Assuming that the condition of full
randomization is fulfilled, the action of the quantum operation
$\varepsilon$ is given by
\begin{equation}\label{eps}
\varepsilon: \rho\rightarrow\varepsilon(\rho)=
\frac{\mathds{1}}{d}
\end{equation}
for every $\rho$, that is, an arbitrary (pure) state is mapped onto a
maximally-mixed state. We recall that an arbitrary density matrix can
be written as
\begin{equation}
\rho=\frac{\mathds{1}}{d}+\sum_{j=1}^{d^2-1}a_jT_j \quad (a_j\in\mathbb{R})\:.
\end{equation}
Since $\varepsilon(\rho)=\mathds{1}/d$ for an arbitrary $\rho$, by
assumption, it follows by linearity and trace-preserving that
$\varepsilon(T_j)=0$, for every $j=1,\ldots,d^2-1$. Hence, the sum in
Eq.~\eqref{avgatefid2} evaluates to $0$, implying that the average
gate fidelity for the mapping in Eq.~\eqref{eps} is given by
\begin{equation}
\bar{F}(\varepsilon,U)=\frac{1}{d}\:.
\label{sat}
\end{equation}
Note that Eq.~\eqref{sat} is applicable only if the randomness in the
control field amplitudes allows the uniform generation of any unitary
contained in $SU(d)$.  For example, in the three-spin ($d=8$) case we
expect the average gate fidelity to saturate at a value of
$0.125$. This seems to be numerically corroborated by
Fig.~\ref{sensitivity_hxhy} and indicates that the system undergoes
full randomization.  Also in a four-spin chain ($d=16$), provided that
full randomization occurs, the found saturation value fits well with
$1/d=0.0625$ (not illustrated here). It should be emphasized that this
regime of full randomization is of no relevance for practical
realizations of control, however, it provides an important consistency
check of our numerical results.

We now address the observation that there exists no universal
saturation value in the case of control by $x$ pulses only, see
Fig.~\ref{sensitivity_hx}. This type of control does not allow
universal quantum computation, i.e., the reachable set of unitary
operators is reduced to a Lie subgroup of $SU(d)$. Since
Eq.~(\ref{avgatefid1}) is only valid if the reachable set equals the
full $SU(d)$, the reasoning leading to Eq.~(\ref{sat}) does not apply
to such a subgroup. The accessible part of the reachable set depends
on the evolution time $t_f$. The control fields of
Fig.~\ref{sensitivity_hx}(b) differ in $t_f$ which may result in
different accessible parts of the reachable sets and, consequently,
different saturation values. The fact that in
Fig.~\ref{sensitivity_hx}(a), where all fields have the same $t_f$,
the average gate fidelity saturates to the same value corroborates
this explanation.

\subsection{Physical interpretation} \label{nmr}

As illustrated in Figs.~\ref{sensitivity_hxhy} and~\ref{sensitivity_hx}, 
the average fidelity is less susceptible to random errors for shorter time 
intervals $T$. This is a special case of a more general physical situation, 
namely a competition between resonance- and relaxation-type behavior~\cite{DattaguptaBook}. 
A typical example is the phenomenon of motional narrowing of the linewidth in NMR
experiments~\cite{CallaghanBook}.

This generic class of phenomena is discussed based on models
for a randomly-interrupted deterministic motion~\cite{DattaguptaBook}: a system is subjected
to a random event ( e.g., switching between magnetic fields $B_{0}$ and
$-B_{0}$ in NMR) with the switching rate $\lambda$, while in between two
random events it evolves deterministically. On the whole, the system
is then evolving piecewise-deterministically and its qualitative
behavior depends on the relative magnitude of the
switching rate $\lambda$ and the average (absolute) external field
amplitude ($B_{0}$ in the example above).  Note that with units chosen
such that $\hbar=1$ both quantities have the same physical dimensions.

In our system, for smaller time intervals $T$ the fidelity in the
presence of random errors is indeed closer to the intrinsic optimal
values. Interestingly, $\lambda=T^{-1}\sim J$ in the optimal
cases, while the average (absolute) control-field amplitudes (that are
independent of $\delta$ because of the symmetry of the probability
distribution) are of the same order of magnitude. This explains the
dependence of $\bar{F}$ on $\delta$: the fidelity varies significantly
with $\delta$ and saturates at a value around $0.125$ in the case of a
completely controllable three-spin chain ($0.0625$ in the four-spin
chain). In contrast to that, for $\lambda$ much bigger than the
average control-amplitudes, the fidelities would remain close to $1$
even for large $\delta$.

\section{Spectral filtering} \label{smoothing}

Our choice of piecewise-constant control fields is, at least partly, a
matter of mathematical convenience. In realistic implementations of
quantum control various constraints may come into play, the most
important ones pertaining to the complexity of the frequency spectrum
of permissible control fields. In control experiments that make use
of an external magnetic field, for instance, such constraints are
related to bandwidth and slew-rate limitations of the magnetic coils
and drivers~\cite{Chaudhury+:07}. Therefore, it is necessary to
subject the optimal control fields to spectral filtering in order to
make contact with experimental realizations~\cite{Jirari+:09}.

\begin{figure}[t!]
\includegraphics[width=0.85\linewidth]{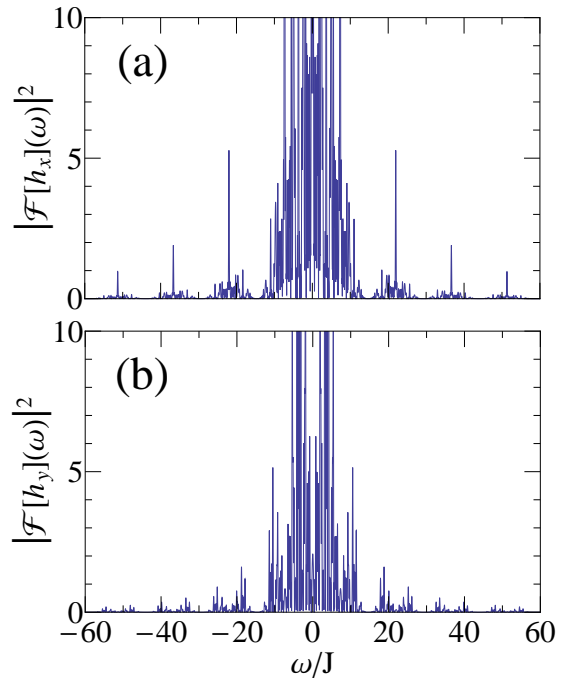}
\caption{\label{power_spectrum}(Color online) The power spectra 
corresponding to the optimal $x$ [(a)] and $y$ [(b)] control fields
shown in Fig.~\ref{optimal_fields}(a).}
\end{figure}

In what follows, we aim at finding control fields that are smoother
than the optimal piecewise-constant ones, at the same time retaining
high fidelities for our target quantum gates. To this end, we put
constraints on the frequency spectra of the control fields $h_j(t)$
($j=x,y$) by means of frequency filter functions. After operating with a filter function
$f(\omega)$ on the Fourier transforms $\mathcal{F}[h_j]$ of the optimal fields, we switch
back to the time domain and calculate the filtered control fields
$\widetilde{h}_j(t)$ via inverse Fourier
transformation~\cite{Werschnik+Gross:07}:
\begin{equation}\label{filtering_procedure}
\widetilde{h}_j(t)=\mathcal{F}^{-1}\big[f(\omega)
\mathcal{F}[h_j](\omega)\big](t) \quad (j=x,y)\:.
\end{equation}
The power spectrum of a typical optimal piecewise-constant control
field is depicted in Fig.~\ref{power_spectrum}.

Based on the obtained smoothened control fields, we calculate the 
corresponding fidelities for different quantum gates, using a product-formula 
approximation (for details, see the Appendix).

\begin{figure}[t!]
\includegraphics[width=0.85\linewidth]{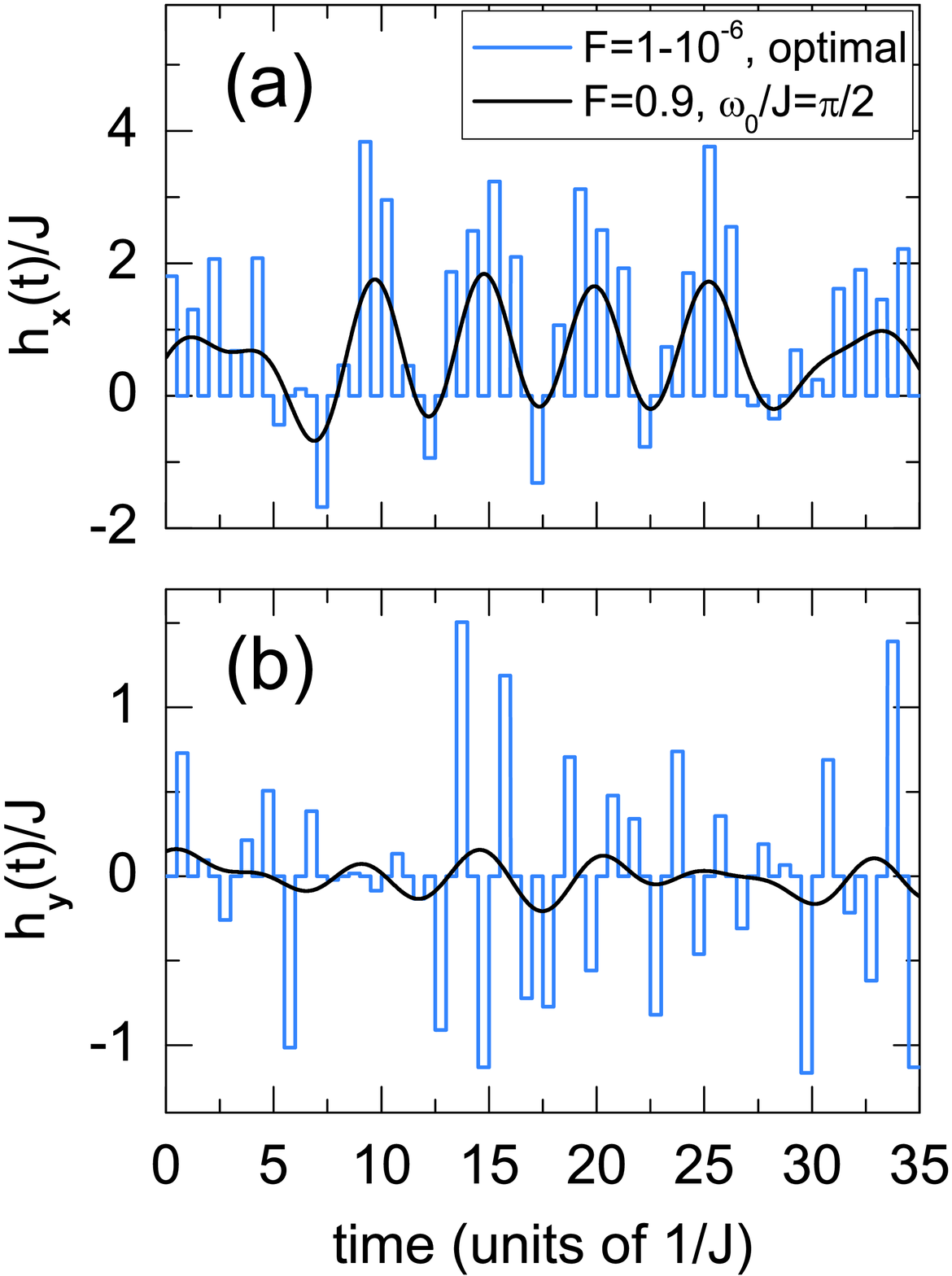}
\caption{\label{Filter11X_low-pass}(Color online) Optimal
  piecewise-constant control field which realizes the $X_3$ gate by
  alternate $x$ [(a)] and $y$ [(b)] pulses, compared to the low-pass
  filtered counterpart (cut-off frequency $\omega_0/J=\pi/2$) with a
  fidelity of $0.9$. The optimal field corresponds to $N_t=70$ and
  $T=0.5$.}
\end{figure}

\begin{figure}[t!]
\includegraphics[width=0.85\linewidth]{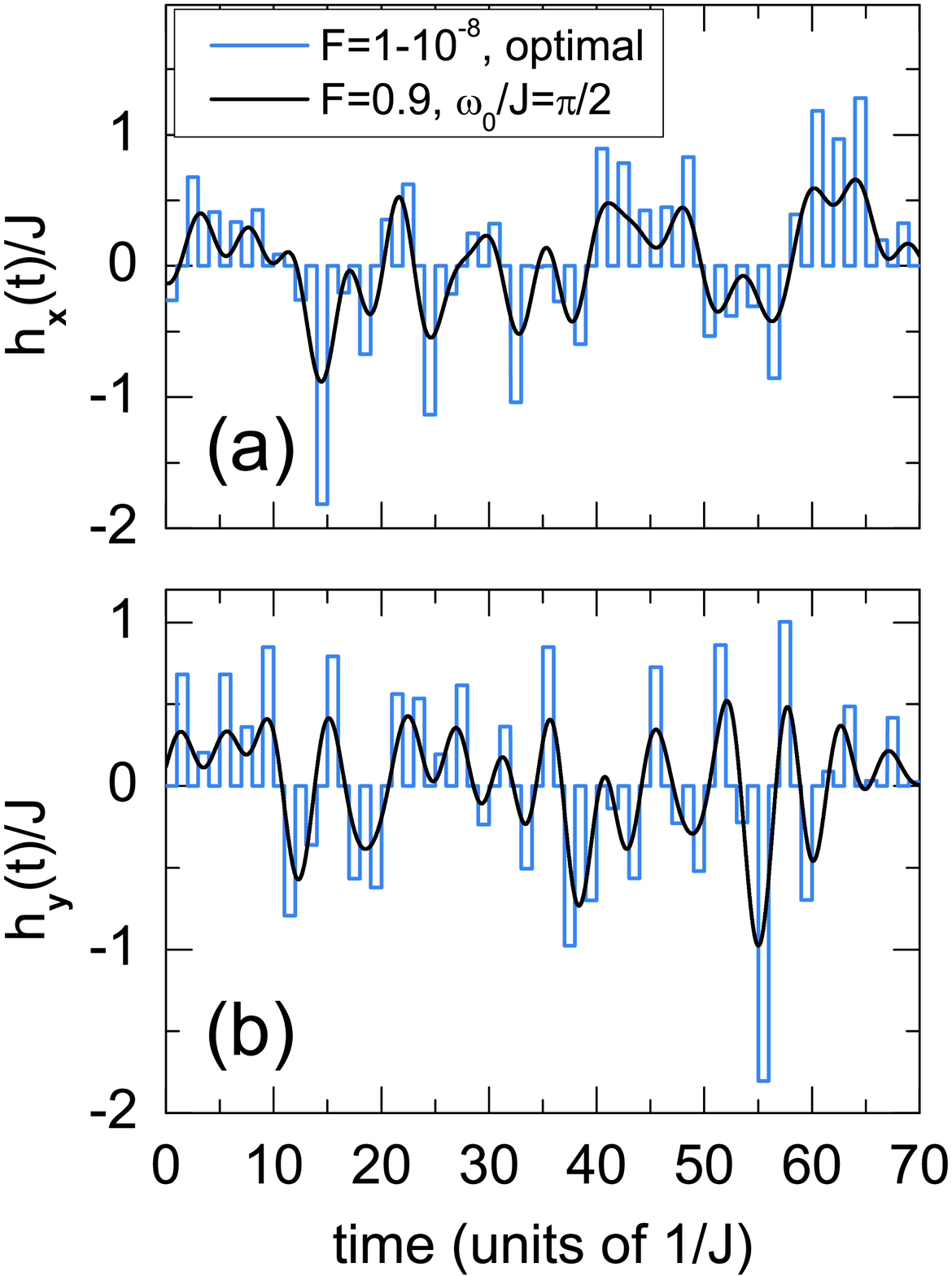}
\caption{\label{Filter1CNOT_low-pass}(Color online) Optimal
  piecewise-constant control field which realizes the
  $\mathrm{CNOT}_3$ gate by alternate $x$ [(a)] and $y$ [(b)] pulses,
  compared to the low-pass filtered counterpart (cut-off frequency
  $\omega_0/J=\pi/2$) with a fidelity of $0.9$. The optimal field
  corresponds to $N_t=70$ and $T=1.0$.}
\end{figure}

We first consider an {\em ideal low-pass} filter which removes
frequencies outside the interval $[-\omega_0,\omega_0]$:
$f(\omega)=\theta(\omega+\omega_0)-\theta(\omega-\omega_0)$.
Using the general prescription in Eq.~\eqref{filtering_procedure}, we obtain
\begin{equation}\label{tildeh}
\begin{split}
\widetilde{h}_x(t)=\frac{1}{\pi}\sum_{n=1}^{N_t/2}
h_{x,n}\big[a_{2n-1}(t)-a_{2n-2}(t)\big]\:,\\
\widetilde{h}_y(t)=\frac{1}{\pi}\sum_{n=1}^{N_t/2}
h_{y,n}\big[a_{2n}(t)-a_{2n-1}(t)\big]\:,
\end{split}
\end{equation}
where $a_{m}(t)\equiv\mathrm{Si}\big[\omega_0(mT-t)\big]$ and
$\mathrm{Si}(x)\equiv\int_0^{x}(\sin{t}/t)dt$. 
Two examples of such smoothened pulses for the $X_3$ and $\mathrm{CNOT}_3$ gates,
corresponding to a fidelity of $0.9$, are shown in
Figs.~\ref{Filter11X_low-pass} and \ref{Filter1CNOT_low-pass}.

\begin{figure}[t!]
\includegraphics[width=0.85\linewidth]{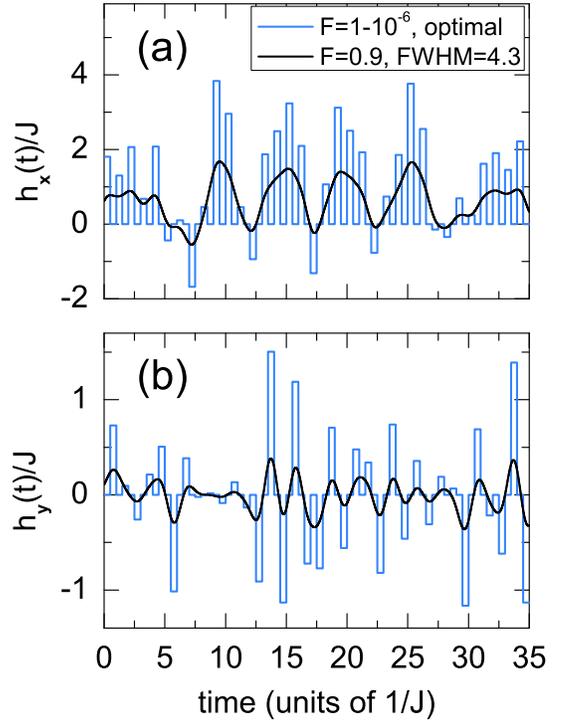}
\caption{\label{Filter11X_gaussian}(Color online) Optimal
  piecewise-constant control field which realizes the $X_3$ gate by
  alternate $x$ [(a)] and $y$ [(b)] pulses, compared to the Gaussian
  filtered counterpart ($\mathrm{FWHM}=4.3$) with a fidelity of
  $0.9$. The optimal control field corresponds to $N_t=70$ and
  $T=0.5$ as in the case of low-pass
  filtering shown in Fig.~\ref{Filter11X_low-pass}.}
\end{figure}

We also consider {\em Gaussian filters} with center frequencies
$\pm\omega_{c}$: 
$f(\omega)=\exp[-\gamma(\omega-\omega_{c})^2]+\exp[-\gamma(\omega+\omega_{c})^2]$,
where $\gamma>0$ determines the Gaussian full-width-at-half-maximum
$\mathrm{FWHM}=2\sqrt{\ln{2}/\gamma}$.  As the power spectra of the
optimal control fields (Fig.~\ref{power_spectrum}) show, the dominant
frequencies are located around $\omega=0$.
Hence we choose $\omega_c=0$ and vary the Gaussian width.
We determine the smoothened control fields by making use of the
identity~\cite{GradshteynRyzhik}
\begin{equation}
\int_{0}^{\infty}e^{-p^2x^2}\frac{\sin{(ax)}}{x}\:dx
=\frac{\pi}{2}\:\mathrm{erf}\left(\frac{a}{2p}\right)\:,
\end{equation}
where $\mathrm{erf}$ is the error function. We obtain 
\begin{equation}
\begin{split}
\widetilde{h}_x(t)=\frac{1}{2}\sum_{n=1}^{N_t/2}
h_{x,n}\big[b_{2n-1}(t)-b_{2n-2}(t)\big]\:,\\
\widetilde{h}_y(t)=\frac{1}{2}\sum_{n=1}^{N_t/2}
h_{y,n}\big[b_{2n}(t)-b_{2n-1}(t)\big]\:,
\end{split}
\end{equation}
where $b_{m}(t)\equiv\mathrm{erf}\left[(mT-t)/(2\sqrt{\gamma})\right]$.
The results obtained through Gaussian filtering resemble the ideal
low-pass ones.  In Fig.~\ref{Filter11X_gaussian} we show the optimal
sequence of controls realizing the $X_3$ gate
which is now smoothened by applying a Gaussian filter, retaining, however,
a fidelity of $0.9$ (compare the corresponding low-pass filtered
pulse in Fig.~\ref{Filter11X_low-pass}). 

\begin{figure}[b!]
\includegraphics[width=0.85\linewidth]{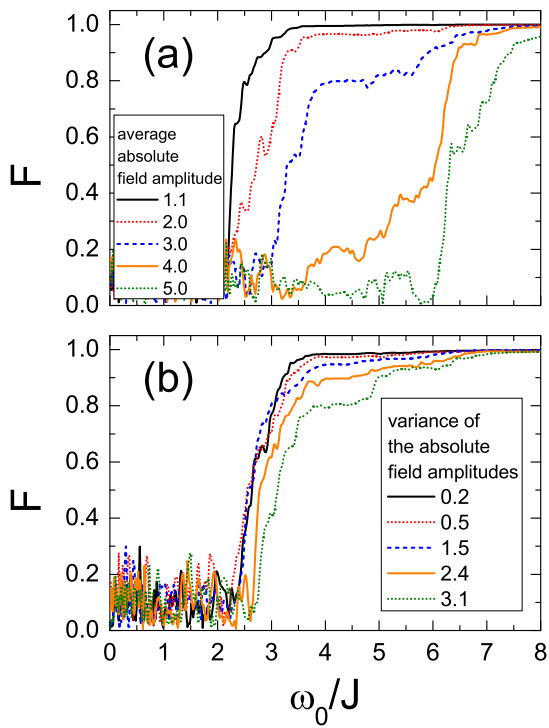}
\caption{\label{Filter11X_low-pass_fidelity}(Color online) Fidelity
  ($F$) versus cut-off frequency ($\omega_0$) for different low-pass
  filtered optimal control fields, all of them implementing the $X_3$
  gate with $70$ pulses of durations $1.5$ alternately applied in the
  $x$ and $y$ directions: (a) The optimal control fields with average
  amplitudes indicated and variances ranging between $1.1$ and $1.3$.
  (b) The optimal control fields with average amplitudes between $1.9$
  and $2.1$, and variances specified in the legend.}
\end{figure}

The behavior of the fidelity versus the cut-off (low-pass) or the
width (Gaussian filtering) turns out to be dependent on the strength
and pulse shape of the optimal control
fields. Figure~\ref{Filter11X_low-pass_fidelity} illustrates this
dependence in the case of ideal low-pass filtering 
which implement the $X_3$ gate. The
curves in Fig.~\ref{Filter11X_low-pass_fidelity}(a) reveal that the
fidelity of weaker optimal control fields is less affected by the
low-pass filtering. The average absolute field amplitude specified in
the legend has been calculated by averaging the absolute values of the
pulse amplitudes over the total control time $t_f$. 
In contrast, the optimal fields corresponding to the fidelity
curves shown in Fig.~\ref{Filter11X_low-pass_fidelity}(b) are
characterized by a similar average absolute amplitude but differ in
the variance of the single absolute pulse amplitudes defined with
respect to the average absolute amplitude. We see that the fidelity
decays faster if the variance is larger, i.e., if the pulse amplitudes
of the field are varying more rapidly.  Analyzing in
Fig.~\ref{Filter11X_low-pass_fidelity} the cut-offs at which the
fidelity decays, we estimate that frequencies of up to at least twice
the average absolute field amplitude are required to retain fidelities
of $0.9$ or larger.

To improve the results from the filtering of optimal control fields,
we tried to iterate the procedure: The filtered fields are discretized and used
as initial guess for the optimization algorithm.  The optimal fields
produced by iteration are then filtered again. In so doing we intended
to generate pulses with high fidelity and even lower frequency than in
the ``ordinary filtering'' discussed before. However, this approach
does not yield significant improvements.

\section{Summary and Conclusions} 
\label{sumconclude}

Recent quantum-control studies have shown that an array of qubits
coupled by nearest-neighbor Heisenberg interactions can be universally
operator-controlled by acting only on the first qubit.  Since these
results only imply the existence of a control sequence but do not
provide a way to construct them, we have investigated the feasibility
of controlling a spin chain with Heisenberg interaction by applying a
time-dependent magnetic field to the first spin in the chain. We have
explicitly determined piecewise-constant control fields for several
quantum gates for three- and four-spin chains. By increasing the
number of control pulses within a fixed total time, fidelities
arbitrarily close to $1$ can be achieved.

We have also studied the sensitivity of the fidelity to random errors
in the control fields.  Our analysis shows that the average fidelity
is less susceptible to random perturbations if the durations of the
single control pulses are reduced. This behavior is related to a
generic class of phenomena exemplified by motional narrowing in NMR.
We have also examined the intrinsic saturation of the average
fidelity, being universal for complete controllability and occurring
if the strength of the acting random field is large enough. Finally,
to make contact with experimental realizations, we have used spectral
filtering to obtain smoothened control pulses retaining fidelities of
$0.9$.

Our study can be extended to include more complicated (e.g.,
continuously varying) control pulses.  Such cases would require the
use of some advanced methods of the optimal control theory, e.g., the
Krotov algorithm, which was shown to be capable of reaching the
quantum speed limit~\cite{Caneva+:09}.

The statement that universal control of a qubit array is possible by
acting on the first qubit is true in principle for arbitrarily long
arrays. However, the complexity of finding the control sequence will
make such a procedure impractical for Heisenberg chains.  We expect
our results to be of practical significance for relatively small
systems of a few qubits, like those realized experimentally right
now. Having fewer control lines will increase the coherence time of
the qubits that are indirectly controlled by their interaction with
the neighboring qubits. In the present paper, we have studied
isotropic Heisenberg interactions, but the procedure also works for
$XXZ$-type interactions.  We therefore expect that our work will
facilitate the control of quantum information devices with relatively
few qubits.

\begin{acknowledgments}
This work was financially supported by EU project SOLID, the EPSRC
grant EP/F043678/1, the Swiss NSF, and the NCCR Nanoscience.
\end{acknowledgments}

\appendix
\section{Fidelity calculation for smoothened control fields}\label{app}

In order to determine the fidelity corresponding to the smoothened
control fields, we have to calculate the time-evolution operator
$\widetilde{U}(t)$ governed by the Hamiltonian
$\widetilde{H}(t)=H_0+\widetilde{h}_x(t)S_{1x}+\widetilde{h}_y(t)S_{1y}$. One
possible approach to this problem is solving the equation of motion
for $\widetilde{U}(t)$, which can be written as a system of $d^2$
coupled first-order differential equations
\begin{equation}\label{odesystem}
i\frac{d}{dt}\:\widetilde{U}_{kl}(t)=\sum_{j=1}^d 
\widetilde{H}_{kj}(t)\widetilde{U}_{jl}(t) \quad (k,l=1,...,d)\:,
\end{equation}
with $\widetilde{U}_{kl}(0)=\delta_{kl}$. While Runge-Kutta-type methods
are commonly utilized for solving such systems, they are not so
convenient in the case at hand where it is essential to preserve the
unitarity of the time-evolution operator. We therefore determine
$\widetilde{U}(t_f)$ using a product-formula approximation which
manifestly preserves unitarity and essentially amounts to a
discretization in time.

We divide each time interval of length $T$ into $m_{\scriptscriptstyle
  T}$ steps of length $\tau\equiv T/m_{\scriptscriptstyle T}$;
the total number of steps is denoted by 
$m_f\equiv N_t m_{\scriptscriptstyle T}$. 
The approximation hinges on the assumption
that during each time step the total Hamiltonian of the system remains
constant, i.e., time-independent. In other words
\begin{equation}
\widetilde{H}(t)=H^{(k)}\equiv H_0+\widetilde{h}_x(k\tau)S_{1x}
+\widetilde{h}_y(k\tau)S_{1y} 
\end{equation}
for $k\tau\leq t<(k+1)\tau$, with $k=0,\ldots,m_{f}-1$. 
Hence the evolution of the system during the $(k+1)$-th interval 
is described by $U^{(k)}(\tau)\equiv e^{-iH^{(k)}\tau}$.
Using the semigroup property of time-evolution operators,
$\widetilde{U}(t_f)$ can be expressed as the product
\begin{equation}\label{Uprod}
\widetilde{U}(t_f)=e^{-iH^{(m_f-1)}\tau}\cdot...\cdot e^{-iH^{(1)}\tau}
\cdot e^{-iH^{(0)}\tau}\:,
\end{equation}
where each of the operators $e^{-iH^{(k)}\tau}$ is found using the
spectral representation (recall Sec.~\ref{pccontrols}).  This
approximation becomes progressively more accurate with decreasing
$\tau$, so that the time-evolution operator can be computed to the
required accuracy.  The unitarity of the time-evolution operator is
preserved by construction for an arbitrary number of time steps, hence
the method is unconditionally stable~\cite{DeRaedt:00}.

\bibliography{QuantumControl}
\bibliographystyle{apsrev}

\end{document}